\documentclass[conference,a4paper]{IEEEtran}

\usepackage{amsmath}
\usepackage{amsfonts}
\usepackage{amssymb}
\usepackage{amsbsy}
\usepackage{graphicx}
\usepackage{times}
\usepackage{default}
\usepackage{cite}
\usepackage{enumerate}
\def\av{\boldsymbol{a}}
\def\ev{\boldsymbol{e}}

\def\uv{\boldsymbol{u}}\def\vv{\boldsymbol{v}}\def\wv{\boldsymbol{w}}\def\xv{\boldsymbol{x}}
\def\yv{\boldsymbol{y}}\def\zv{\boldsymbol{z}}\def\0v{\small{\boldsymbol{0}}}\def\1v{\boldsymbol{1}}

\def\am{\boldsymbol{A}}\def\bm{\boldsymbol{B}}
\def\hm{\boldsymbol{H}}

\def\arm{\boldsymbol{\mathsf{A}}}

\def\arv{\boldsymbol{\mathsf{a}}}

\def\xrv{\boldsymbol{\mathsf{x}}}
\def\yrv{\boldsymbol{\mathsf{y}}}\def\zrv{\boldsymbol{\mathsf{z}}}

\def\as{\mathcal{A}}\def\bs{\mathcal{B}}\def\cs{\mathcal{C}}\def\ds{\mathcal{D}}

\def\ss{\mathcal{S}}\def\ts{\mathcal{T}}
\def\us{\mathcal{U}}

\def\IN{\mathbb{N}}

\def\IR{\mathbb{R}}

\def\IrX{\mathbb{\mathsf{X}}}
\def\IrY{\mathbb{\mathsf{Y}}}
\def\IrZ{\mathbb{\mathsf{Z}}}

\def\pr{\mathbb{P}}

\def\rank{\operatorname{rank}}

\def\spt{\operatorname{spt}}
\def\ee{\mathbb{E}}
\def\leb{\operatorname{Leb}}

\def\dim{\operatorname{dim}}

\def\MIL{\underline{\dim}_\text{B}}
\def\MIU{\overline{\dim}_\text{B}}
\def\MILU{{\dim}_\text{B}}

\def\ba#1\ea{\begin{align*}#1\end{align*}}	%Shortcut for align-environment
\def\ban#1\ean{\begin{align}#1\end{align}}	%Shortcut for align-environment, numbered
\def\bac#1\eac{\vspace{\abovedisplayskip}{\par\centering$\begin{aligned}#1\end{aligned}$\par}\addvspace{\belowdisplayskip}}	% the same, centered

\newcommand{\lefto}{\mathopen{}\left}

\newtheorem{theorem}{Theorem}

\newtheorem{lemma}{Lemma}
\newtheorem{definition}{Definition}
\newtheorem{proposition}{Proposition}
\newtheorem{corollary}{Corollary}
\newtheorem{remark}{Remark}%[section]

\title{Almost Lossless Analog Signal Separation}

\author{
\IEEEauthorblockN{David Stotz$^1$, Erwin Riegler$^{2}$,  and  Helmut B\"olcskei$^1$}\\
\IEEEauthorblockA{
$^1$ETH Zurich,  Switzerland\\
$^2$Vienna University of Technology, Austria
}}

\begin{document}
\maketitle
\begin{abstract}
%THIS PAPER IS ELIGIBLE FOR THE STUDENT PAPER AWARD. 
We propose an information-theoretic framework for analog signal separation. Specifically, we consider the problem of recovering two analog signals from a noiseless sum of linear measurements of the signals. Our framework is inspired by the groundbreaking work of Wu and Verd\'u (2010) on almost lossless analog compression. The main results of the present paper are a general achievability bound for the compression rate in the analog signal separation problem, an exact expression for the optimal compression rate in the case of signals that have mixed  discrete-continuous distributions, and a new technique for showing that the intersection of generic subspaces with subsets of sufficiently small Minkowski dimension is empty.
This technique can also be applied to obtain a simplified proof of a key result in Wu and Verd\'u (2010). 

\end{abstract}

\section{Introduction}
We consider the following signal separation problem: Reconstruct the vectors $\yv$ and $\zv$ from the noiseless observation 
\begin{align}\label{eq:problem1} 
\wv=\am\yv+\bm\zv
\end{align}  
where $\am$ and $\bm$ are (measurement) matrices. As detailed in \cite[Sec.~1]{SKP12} this  
problem 
has numerous applications such as inpainting, super-resolution, and the recovery of clipped signals and of signals that are corrupted by impulse noise or narrowband interference. 
 
The  sparse signal recovery literature \cite{Li12, WM10, DMM09, DK10, Tro08, CT12, PBS13, SKP12, DH01, DS89} provides separation guarantees under sparsity constraints on the vectors $\yv$ and $\zv$. The sparsity  
thresholds in \cite{SKP12,DK10,DH01,DS89} are functions of the coherence parameters \cite{SKP12} of the matrices $\am$ and $\bm$ and hold for \emph{all}  $\yv$ and $\zv$ meeting these thresholds, but suffer from the square-root bottleneck \cite{Tro08}. 
For random signals, the asymptotic results in \cite{Li12, WM10, PBS13} overcome the square-root bottleneck, but hold ``only'' with overwhelming probability.  
For $\bm$ the identity matrix and $\am$ a random orthogonal matrix it is shown in \cite{CT12}  that the probability of failure of a certain reconstruction procedure decays exponentially in the dimension of the ambient space.

\paragraph*{Contributions}
Inspired by the recent work of Wu and  Verd\'u \cite{WV10}, we derive asymptotic recovery results for the analog signal separation problem with  the vectors $\yv$ and $\zv$ random, possibly dependent, and of general distributions. Our results hold for deterministic  $\bm$ and for almost all (a.a.) matrices $\am$, but do not depend on coherence parameters.  
However, since we assume $\yv$ and $\zv$ to be random, the statements are in terms of probability of separation error with respect to the source distributions, and hence do not provide worst-case guarantees like the coherence-based results in \cite{SKP12,DK10,DH01,DS89}. 

Specifically, we study the asymptotic setting $\ell, n \to \infty$ where the vectors $\yv\in\IR^{n-\ell}$ and $\zv\in\IR^{\ell}$ are realizations of  random processes; for each $n$, we let $\ell=\lfloor \lambda n\rfloor$ and $k=\lfloor R n\rfloor$ for parameters $R,\lambda\in [0,1]$ and measurement matrices $\am\in\IR^{k\times (n-\ell)}$ and $\bm\in \IR^{k\times \ell}$, with $k\geqslant \ell$.  We refer to the parameter $R$ as the compression rate as it equals (approximately) the ratio between the number of measurements and the total number of parameters in $\yv$ and $\zv$.
In Theorem~\ref{th2}, we show that for each (deterministic) full-rank matrix $\bm$, recovering $\yv$ and $\zv$ from the measurement $\wv$ is possible with arbitrarily small probability of separation error for a.a.\ matrices $\am$, provided that $n$ is sufficiently large and the compression rate $R$ is larger than the Minkowski dimension compression rate (see\ Definition~\ref{defminkowskirate}) of the concatenated random vector $[\yv^{\operatorname{T}} \:\: \zv^{\operatorname{T}}]^{\operatorname{T}}$. 
Since the technique used to prove \linebreak the related result \cite[Thm.~18]{WV10} in the context of almost lossless analog compression can not be adapted to our setting, we develop a new proof method. The foundation of our approach, inspired by \cite{SYC91}, is a new technique for showing that the intersection of generic subspaces with subsets of sufficiently small Minkowski dimension is empty (Proposition~\ref{prop:inj}). A novel  concentration of measure result, developed in Lemma~\ref{lemma2}, turns out to be an essential ingredient of this technique. Applying our method to the setting in \cite{WV10} leads to a significant simplification of the proof of \cite[Thm.~18, 1)]{WV10}.

For $\yv$ and $\zv$ mixed discrete-continuously distributed with mixing parameters $\rho_1$ and  $\rho_2$, respectively, we show that the Minkowski dimension compression rate can be evaluated explicitly to 
\begin{align}\label{eq:threshold}
(1-\lambda)\rho_1+\lambda\rho_2.
\end{align}
What is more, this threshold is tight in the sense that there is a converse if the compression rate $R$ is smaller than \eqref{eq:threshold}.

\paragraph*{Notation}
For a relation $\# \in\{<,>,\leqslant,\geqslant,=,\neq,\in,\linebreak \notin\}$, we write 
$f(n)\ \overset{\textbf{.}}{\#}\ g(n)$ if there exists an $N\in \IN$ such that $f(n)\ \# \ g(n)$ holds for all $n>N$. 
$\leb^n$ denotes the $n$-dimensional Lebesgue measure and $\bs^{\otimes n}$ the Borel $\sigma$-algebra on $\IR^n$. We write $\|\,\cdot \,\|$ for  the $\ell_2$-norm on $\IR^n$.    
Matrices are denoted by capital boldface and vectors by lowercase boldface letters. $B^n(\xv,\varepsilon)$ is the ball centered at $\xv\in\IR^n$ of radius $\varepsilon$ with respect to  $\|\,\cdot\, \|$, and   its  volume is  $\alpha(n,\varepsilon)=\leb^n(B^n(\xv,\varepsilon))$.  The closure of a set   $\us\subset\mathbb R^n$ is  denoted by $\overline{\us}$. We use sans-serif letters, e.g.\ $\xrv$, for random quantities and roman letters, e.g. $\xv$, for deterministic quantities. For a random variable $\IrX$ or a random vector $\xrv$, $\mu_{\IrX}$ and $\mu_{\xrv}$ denote the respective distribution. We write $\1v_{\IrX\in\as}$ for the characteristic function associated with the event $\IrX \in \as$.  

\section{Main result}
We start by noting that \eqref{eq:problem1} can be rewritten as 
\ba 		\wv = [\am\:\: \bm] \begin{bmatrix} \yv \\ \zv \end{bmatrix}\ea
which shows that, formally, the separation problem we consider can be cast as an almost lossless analog  compression problem  \cite{WV10} with 
measurement matrix $\hm=[\am\:\: \bm]$ and random source vector $[\yrv^{\operatorname{T}}\; \zrv^{\operatorname{T}}]^{\operatorname{T}}$, where $\yrv$ and $\zrv$ are possibly dependent. As we shall see below in Remark \ref{remgehtnicht}, the results in \cite{WV10} can, however, not be applied to our setting,  as $\bm$ is deterministic here, whereas the results in \cite{WV10} hold for a.a.\ matrices $\hm$.

\begin{definition}\label{definitionX}
Let $0\leqslant \lambda \leqslant 1$.  Suppose that $(\IrY_i)_{i\in\IN}$ and $(\IrZ_i)_{ i\in\IN}$ are stochastic processes on $(\IR^\IN,\bs^{\otimes\IN})$. 
Then, for $n\in\IN$, the source vector $\xrv$ of length $n$ is given by $\xrv=[\IrX_1 \, \dots\, \IrX_n]^{\operatorname{T}}$ with 
\begin{align*}
\IrX_i &= \IrY_i,\quad \text{for}\ i\in\{1,\dots,n-\ell\}\\
\IrX_{n-l+i} &= \IrZ_i,\quad \text{for}\ i\in\{1,\dots,\ell\}
\end{align*}
where $\ell=\lfloor \lambda n \rfloor$.
\end{definition}

The rate Definitions~\ref{defsourcerate} and \ref{defminkowskirate} below are adapted from the corresponding definitions in \cite{WV10}.
\begin{definition}(Analog compression - \emph{linear} measurements/ \emph{measurable} separator)\label{defsourcerate}.
For $\xrv$ as in Definition \ref{definitionX} and $\varepsilon >0$, an $(n,k)$ code consists of 
\begin{enumerate}[(i)]
\item linear measurements $[\am\:\: \bm]:\IR^{n-\ell}\times \IR^\ell \to \IR^k$;
\item a separator $g:\IR^k\to \IR^{n-\ell}\times \IR^\ell$ that is measurable with respect to $\bs^{\otimes k}$ and  $\bs^{\otimes n}$.
\end{enumerate} 
We call $R$ with $0\leqslant R\leqslant 1$ an $\varepsilon$-achievable  rate if there exists a sequence of $(n,\lfloor Rn \rfloor)$ codes such that 
\begin{align*}
\pr[g([\am\:\: \bm]\xrv)\neq \xrv]\overset{\textbf{.}}{<}\varepsilon.
\end{align*}
We define the optimal linear compression rate $R_\text{L}(\varepsilon)$ as the infimum over all $\varepsilon$-achievable rates. Here, the name ``linear'' reflects the restriction to linear measurements, employed throughout the paper.
\end{definition}

Next, we define the Minkowski dimension. This quantity is sometimes also referred to as box-counting dimension, which is the origin for the subscript $\text{B}$ in the notation $\MILU(\cdot)$ used below.
\begin{definition}(Minkowski dimension, \cite{Fal04}).\label{defminwowskidim}
Let $\ss$ be a nonempty bounded set in  $\IR^n$.  Define the lower and upper Minkowski dimension of $\ss$ as 
\begin{align*}
\MIL(\ss)&=\liminf_{\varepsilon\to 0}\frac{\log N_\ss(\varepsilon)}{\log\frac{1}{\varepsilon}}\\
\MIU(\ss)&=\limsup_{\varepsilon\to 0}\frac{\log N_\ss(\varepsilon)}{\log\frac{1}{\varepsilon}}
\end{align*}
where $N_\ss(\varepsilon)$ is the covering number of $\ss$ given by 
\begin{align*}
N_\ss(\varepsilon)&=\min\Big\{m \in\IN\mid \ss\subset \bigcup_{i\in\{1,\dots,m\}} B^n(\xv_i,\varepsilon),\ \xv_i\in \IR^n\Big\}.
\end{align*}
If $\MIL(\ss)=\MIU(\ss)$, we simply write $\MILU(\ss)$. 
\end{definition}
\begin{definition}(Minkowski dimension compression rate).\label{defminkowskirate}
For $\xrv$ from Definition \ref{definitionX} and $\varepsilon >0$, we define the Minkowski dimension compression rate as  
\begin{align}
R_\text{B}(\varepsilon)&=\limsup_{n\to\infty} a_n(\varepsilon),\quad\text{where} \nonumber \\
a_n(\varepsilon)&=\inf\Big\{\frac{\MIL(\ss)}{n}   \;\Big\vert\;  \ss \subset\IR^n,\ \pr[\xrv\in\ss]\ \geqslant 1-\varepsilon\Big\}. \label{eq:appsupp}
\end{align} 
\end{definition}

\begin{remark}
Note that in \eqref{eq:appsupp} the infimum is taken with respect to the \emph{lower} Minkowski dimension, whereas the corresponding definition in \cite{WV10} is based on the \emph{upper} Minkowski dimension. Our main result, Theorem~\ref{th2} below, when specialized to the setting of \cite{WV10}, i.e., $\lambda=0$, therefore constitutes an improvement of the general achievability result in \cite{WV10}. 
\end{remark}  

The following theorem states that for \emph{every} full-rank matrix $\bm\in\IR^{k\times \ell}$, with $k\geqslant \ell$, every rate $R$ with $R>\! R_\text{B}(\varepsilon)$ is $\varepsilon$-achievable for a.a. $\am$. 
\begin{theorem}\label{th2}
Let $\xrv$ be as in Definition \ref{definitionX} with $\varepsilon >0$ and let  $R>\! R_\text{B}(\varepsilon)$. Then, for every full-rank matrix $\bm\in\IR^{k\times \ell}$, with $k\geqslant \ell$,  and for   a.a. (with respect to $\leb^{k(n-\ell)}$) matrices $\am\in\IR^{k\times {(n-\ell)}}$, where $k=\lfloor Rn \rfloor$, there exists a measurable separator $g$ such that
\begin{align}\label{eq:resultmain}
\pr [g([\am\:\: \bm]\xrv)\neq\xrv] \overset{\textbf{.}}{<} \varepsilon .
\end{align}

\end{theorem}
\proof
See Section \ref{proofTh1}.
\endproof
\begin{remark}
The proof of Theorem \ref{th2} reveals that the minimum $N\in\IN$ for \eqref{eq:resultmain} to hold for all $n>N$ depends only on the distribution of $\xrv$ and is independent of the matrices $\am$ and $\bm$.
\end{remark}

\begin{remark}\label{remgehtnicht}
In \cite[Thm.~18, 1)]{WV10} it was shown that every rate $R$ with $R>\! R_\text{B}(\varepsilon)$ is $\varepsilon$-achievable in almost lossless analog compression for a.a.\ measurement matrices $\hm\in\IR^{k\times n}$. This result is generalized in Theorem~\ref{th2} above to hold for $\hm=[\am\:\: \bm]$ for a given full-rank matrix $\bm$, with $k\geqslant \ell$, for a.a. matrices $\am\in\IR^{k\times (n-\ell)}$. Since in a concrete separation problem we often encounter a particular structure for $\bm$, for example a certain dictionary under which the corresponding signal is sparse, it is important to have the statement hold for \textit{all} matrices $\bm$, instead of only for a.a.\ $\hm=[\am\:\:  \bm]$. 
The proof of \cite[Thm.~18, 1)]{WV10} relies on intricate properties of measures on Grassmanian manifolds that are invariant under the action of the orthogonal group. These arguments can not be applied to our setting as the overall measurement matrix $\hm=[\am\:\: \bm]$ has a deterministic $k\times \ell$ block $\bm$. 
This forced\linebreak us to find an alternative proof, which is based on two key elements, a concentration of measure result stated in Lemma~\ref{lemma2}, and a dimension counting argument provided in Proposition~\ref{prop:inj}.  The dimension counting argument says that the $(n-k)$-dimensional nullspace of $\hm$ and the approximate support set $\ss$ in \eqref{eq:appsupp} of the source vector $\xrv$ will not intersect, if the Minkowski dimension of $\ss$ is smaller than $k$. Underlying this argument is the basic idea that two objects whose dimensions do not add up to at least the dimension of their ambient space, in general, do not intersect. 
Our proof strategy also applies to the compression problem \cite{WV10} and leads to a significant simplification of the proof of \cite[Thm.~18, 1)]{WV10}, as detailed in Section~\ref{sec:simple}. 
\end{remark}

\section{Mixed discrete-continuous distributions}

In order to establish the connection to the traditional sparse signal separation problem considered, e.g., in \cite{Li12, WM10, DK10, CT12, PBS13, SKP12, DH01, DS89}, we next consider sources $\xrv$ with independent components, where each component  of the constituent processes  $(\IrY_i)_{ i\in\IN}$ and $(\IrZ_i)_{ i\in\IN}$ has a mixed discrete-continuous  distribution, with possibly different mixture parameters for $(\IrY_i)$ and  $(\IrZ_i)$.

\begin{definition}\label{definitionM}
We say that $\xrv$ from Definition \ref{definitionX} has a mixed discrete-continuous distribution if 
for each $n\in\IN$ the random variables $\IrX_i$ for $i\in\{1,\dots,n\}$ are independent and distributed according to  
\begin{align}
\mu_{\IrX_i}=
\begin{cases}
(1-\rho_1)\mu_{\text{d}_1}+\rho_1\mu_{\text{c}_1},\quad i\in\{1,\dots, n-\ell\}\\
(1-\rho_2)\mu_{\text{d}_2}+\rho_2\mu_{\text{c}_2},\quad i\in\{n-\ell+1,\dots, n\} ,
\end{cases}
\end{align}
where $0\leqslant \rho_i\leqslant 1$, the $\mu_{\text{c}_i}$ are distributions on $(\IR,\bs)$, absolutely continuous with respect to Lebesgue measure, and the $\mu_{\text{d}_i}$ are discrete distributions.
\end{definition}

\begin{lemma}\label{lemmax}
Suppose that $\xrv$ is distributed according to Definition \ref{definitionM}. Then 
\begin{align}\label{eq:result1}
R_{\text{B}}(\varepsilon) = \lambda\rho_1 + (1-\lambda)\rho_2  
\end{align}
for all $\varepsilon$ satisfying  $0<\varepsilon <1$.
\end{lemma}
\proof  The proof is largely similar to the proof of \cite[Thm. 15]{WV10}. A sketch of the part that is different is provided in Section \ref{prooflemmax}. \endproof

Theorem~\ref{th2} shows that the optimal linear compression rate  $R_{\text{L}}(\varepsilon)$ is lower-bounded by the Minkowski dimension compression rate $R_{\text{B}}(\varepsilon)$. In the mixed discrete-continuous case we can strengthen this result through the following converse.

\begin{lemma}\label{lemma3}
Suppose that $\xrv$ is distributed according to Definition \ref{definitionM} and let $\varepsilon>0$ and  $R>\! R_\text{B}(\varepsilon)$. Then, for each full-rank matrix $\bm\in\IR^{k\times \ell}$, with $k\geqslant \ell$, and for Lebesgue a.a. (with respect to $\leb^{k(n-\ell)}$) matrices $\am\in\IR^{k\times {(n-\ell)}}$,  where $k=\lfloor Rn \rfloor$, there exists a measurable separator $g$ such that  
\begin{align}\label{eq:errorsepa}
\pr [g([\am\:\: \bm]\xrv)\neq\xrv] \overset{\textbf{.}}{<} \varepsilon.
\end{align}
Moreover, for every $\varepsilon$ with $0<\varepsilon<1$, $R\geqslant R_{\text{B}}(\varepsilon)$ is also a necessary condition for \eqref{eq:errorsepa} to hold, i.e., $R_{\text{L}}(\varepsilon)=R_{\text{B}}(\varepsilon)$.
\end{lemma}
\proof \noindent\emph{Achievability:} 
Follows from Theorem \ref{th2}.

\noindent\emph{Converse:} In the same spirit as the proof of the converse part of \cite[Thm. 6]{WV10}.
\endproof

Finally, we combine Lemmata~\ref{lemmax} and \ref{lemma3} to get an analytical expression for the optimal linear compression rate.

\begin{corollary}\label{sep}
Suppose that $\xrv$ has a mixed discrete-continuous distribution accodring to Definition \ref{definitionM} and let $0<\varepsilon<1$.  
Then, we have
\begin{align}\label{eq:rate}
R_{\text{L}}(\varepsilon)&=(1-\lambda)\rho_1+\lambda\rho_2 .
\end{align}
\end{corollary}

Corollary~\ref{sep} essentially states that the optimal linear compression rate is determined by the fraction of continuously distributed components in the concatenated source vector.
Interestingly, $R_{\text{L}}(\varepsilon)$ does not depend on 
 coherence quantities of the measurement matrices $\am$ and $\bm$, which usually arise in recovery thresholds in the sparse signal separation problem, see, e.g., \cite{SKP12, PBS13}. In this respect, under the rate constraint $R>\! R_\text{B}(\varepsilon)$, a.a.\ matrices $\am$ are ``incoherent'' to a given matrix $\bm$. 
When the distribution of one of the signals is purely discrete, the optimal linear compression rate is determined solely by the distribution of the other signal. Finally, if the dimension of one of the signals is much larger than the dimension of the other, i.e., $\lambda \approx 0$ or $\lambda\approx 1$, then the characteristics of the higher-dimensional signal dominate the threshold in Corollary~\ref{sep}.

\section{Technical results}
In this section, we collect the main technical results referred to earlier in the paper. These results are important ingredients of the proof of  Theorem~\ref{th2}, detailed in Section~\ref{proofTh1}, and the simplification of the proof of \cite[Thm.~18, 1)]{WV10}, described in Section~\ref{sec:simple}. First, we present a concentration result that bounds the  probability that the norm of the image of a  deterministic vector under a random affine mapping is small.

\begin{lemma}\label{lemma2}
Let $\arm=[\arv_1\, \dots \, \arv_k]^{\operatorname{T}}$ be a random matrix in $\mathbb R^{k\times n}$ where the $\arv_i$ are i.i.d.\ uniform on  the set $B^n(\0v,r)$. Then, for each $ \uv\in\IR^{n}\! \setminus\! \{\0v\}$, each $\vv\in\IR^{k}$, and $\delta>0$, we have
\begin{align*}
\pr[\|\arm\uv+\vv\|<\delta] \leqslant  C(n,k,r)\frac{\delta^k}{\|\uv\|^k},
\end{align*}
where $C(n,k,r)$ is a constant that depends on $n$, $k$, and $r$ only.
\end{lemma}
\proof
\begin{align*}
&\alpha(n,r)^{k}\,\pr[\|\arm\uv+\vv\|<\delta]\\
&=\leb^{kn}\{\am\in B^n(\0v,r)\times \ldots \times B^n(\0v,r)\mid \|\am\uv+\vv\|<\delta\}\\
%&\leqslant \prod_{i=1^k}\alpha(n-l,M)^{-k}\leb^{n-l}\{\av_i\in B(0,M)\mid \|\av^{\operatorname{T}}\uv_1+v_i\|<\delta\} \\
&\leqslant \prod_{i=1}^k\leb^{n}\{\av_i\in B^{n}(\0v,r)\mid |\av^{\operatorname{T}}_i\uv+v_i|<\delta\}\\
%&= \prod_{i=1}^k\leb^{n-l}\Big\{\av_i\in B^{n-l}(0,M)\mid \Big|\av^{\operatorname{T}}_i\frac{\uv}{\|\uv\|}+\frac{v_i}{\|\uv\|}\Big|<\frac{\delta}{\|\uv\|}\Big\}\\
&\overset{(a)}{=}\prod_{i=1}^k\leb^{n}\lefto\{\av_i\in B^{n}(\0v,r)\mid \Big|\av^{\operatorname{T}}_i\ev_1+\frac{v_i}{\|\uv\|}\Big|<\frac{\delta}{\|\uv\|}\right \} \\ 
%&\overset{(b)}{\leqslant} (2M)^{k(n-l-1)}\prod_{i=1}^k\leb^{1}\Big\{a_i\in B^{1}(0,M)\mid \Big|a_i+\frac{v_i}{\|\uv\|}\Big |<\frac{\delta}{\|\uv\|}\Big\}\\
&\overset{(b)}{\leqslant} (2r)^{k(n-1)}\prod_{i=1}^k\leb^{1}\lefto\{a_i\in\IR \mid \Big|a_i +\frac{v_i}{\|\uv\|}\Big|<\frac{\delta}{\|\uv\|}\right \}\\
&=  \frac{(2r)^{k(n-1)}(2\delta)^k}{\|\uv\|^k} ,
\end{align*}
where $(a)$ follows from the fact that $\leb^{n}$ is invariant under rotations and we consider a rotation that takes $\uv/\|\uv \|$ into $\ev_1=[1\, 0 \,\dots \,0]^{\operatorname{T}}$, and  in 
$(b)$ we denote by $a_i$ the first component of the vector $\av_i$  
and use the fact that the magnitudes of the remaining components of $\av_i$ are less than or equal to $r$. 
\endproof

\begin{proposition}\label{prop:inj}
Let $\mathcal S\subset \mathbb R^n$ be such that $d:=\underline{\dim}_{\text{B}}(\mathcal S)<k$. Then 
\ban 	\{ \uv \in \mathcal S\!\setminus\! \{\0v\} \mid \am \uv = \0v\} = \emptyset , 	\label{eq:empty} \ean
for Lebesgue a.a.\ $\am \in \mathbb R^{k\times n}$.
\end{proposition}
\proof
Suppose that $\arm$ is distributed as specified in Lemma~\ref{lemma2}. In order to show that the Lebesgue measure of matrices $\am$ for which \eqref{eq:empty} does not hold is zero, it suffices to prove  that 
\begin{align} \mathbb P[\exists \uv \in \mathcal S\! \setminus\! \{\0v\} : \arm \uv = \0v]= 0, \label{eq:suffices}\end{align}
for $r>0$. We employ a union bound argument  to lower-bound the norm of vectors in $\ss\!\setminus\! \{\0v\}$:
\ban 	\mathbb P[\exists \uv \in \mathcal S&\! \setminus\! \{0\} : \arm \uv = \0v]\nonumber \\ &\leqslant \sum_{j=1}^\infty \mathbb P  [\exists \uv \in \mathcal S\! \setminus\! B^n  (\0v,1/j ) : \arm \uv = \0v]	.	\label{eq:inf}	\ean
This allows us to conclude that it suffices to prove \eqref{eq:suffices} for sets $\mathcal S'\subset \mathcal S$ with $\min \{ \|\uv\| \mid \uv\in\overline{\mathcal S'} \}  >0$, as this would show that each term in  the series in \eqref{eq:inf} is zero.
Using the definition of the Minkowski dimension (Definition~\ref{defminwowskidim}) and the fact that $\mathcal S'\subset \mathcal S$ implies  
$d':=\underline{\dim}_{\text{B}}(\mathcal S')\leqslant d$, we can find a sequence  $\varepsilon_m$ tending to zero such that 
\ba 	\frac{\log \mathcal N_{\mathcal S'}(\varepsilon_m)}{\log \frac{1}{\varepsilon_m}} \xrightarrow{m\to\infty}	d'.	 	\ea
  Let  $\xv_1,\ldots , \xv_{\mathcal N_{\mathcal S'}(\varepsilon_m)}$ be the centers of the balls of radius $\varepsilon_m$ that cover $\mathcal S'$ (cf. Definition~\ref{defminwowskidim}). Since $\min \{ \|\uv\| \mid \uv\in\overline{\mathcal S'} \}>0$, we can assume $m$ to be sufficiently large for  $\min \{ \|\xv_i\|  \mid 1\leqslant i \leqslant \mathcal N_{\mathcal S'}(\varepsilon_m) \}  >0$ to hold. As the norm of each row of $\arm$ is bounded, all realizations of $\arm$ have a common Lipschitz constant, say $L$. Putting things together, we find that 
\ba  	\mathbb P[\exists \uv &\in \mathcal S': \arm \uv = 0 ] \\ &\overset{(a)}{\leqslant} \sum_{i=1}^{\mathcal N_{\mathcal S'}(\varepsilon_m)} \mathbb P[\exists \uv \in B^n(\xv_i, \varepsilon_m): \arm \uv = \0v]\\  & \leqslant \sum_{i=1}^{\mathcal N_{\mathcal S'}(\varepsilon_m)} \mathbb P[\exists \uv \in B^n(\xv_i, \varepsilon_m): \|\arm \uv \|<\varepsilon_m ]\\ & \overset{(b)}{\leqslant} \sum_{i=1}^{\mathcal N_{\mathcal S'}(\varepsilon_m)} \mathbb P[\|\arm \xv_i \|<(L+1)\varepsilon_m ]\\ &\overset{(c)}{\leqslant} C(n,k,r,L) \, \mathcal N_{\mathcal S'}(\varepsilon_m) \, \varepsilon_m^k	\underset{(d)}{\xrightarrow{m\to\infty}}	0, \ea
where $(a)$ follows from a union bound argument, $(b)$ is a consequence of $\|\arm \xv_i \| \leqslant  \|\arm (\xv_i-\uv )\| + \|\arm \uv \|\leqslant L \varepsilon_m+ \|\arm \uv \|$, $(c)$ is by application of Lemma~\ref{lemma2}, and $(d)$ is a consequence of 
\ba 	\frac{\log \mathcal N_{\mathcal S'}(\varepsilon_m)}{\log \frac{1}{\varepsilon_m}}-k=\frac{\log\lefto ( \mathcal N_{\mathcal S'}(\varepsilon_m)\varepsilon_m^k\right)}{\log \frac{1}{\varepsilon_m}} \xrightarrow{m\to\infty}	d'-k<0		.\ea
 We have therefore shown that $\mathbb P[\exists \uv \in \mathcal S': \arm \uv = 0 ]=0$.
\endproof

\begin{remark}
The result in Proposition~\ref{prop:inj} is very intuitive as it says that a generic $(n-k)$-dimensional subspace will intersect a $d$-dimensional object with $d<k$ at most trivially.  A statement similar to   Proposition~\ref{prop:inj} was proven in  \cite[Lem.~4.3]{SYC91}. The result in \cite[Lem.~4.3]{SYC91} applies to linear combinations of Lipschitz mappings, and also gives an upper bound on the lower Minkowski dimension of the set on the left hand side of \eqref{eq:empty} when $d\geqslant k$. The proof of \cite[Lem.~4.3]{SYC91} is based on the singular-value decomposition of $\am$. Our proof above is more direct, but applies to $d<k$ only, the case relevant here.
\end{remark}

Next, we provide a generalization of Proposition~\ref{prop:inj}, which will be needed in the proof of Theorem~\ref{th2}. 

\begin{proposition}\label{prop:inj2}
Let $\mathcal S\subset \mathbb R^n$ be such that $d:=\underline{\dim}_{\text{B}}(\mathcal S)<k$, and let $\bm\in\mathbb R^{k\times \ell}$ be a  matrix with $\rank(\bm)=\ell$. Then, 
\ba 	\{ \uv \in \mathcal S\!\setminus\! \{\0v\} \mid [\am\:\: \bm] \uv = \0v\} = \emptyset ,\ea
for Lebesgue a.a.\ $\am \in \mathbb R^{k\times (n-\ell)}$.
\end{proposition}

The proof of Proposition~\ref{prop:inj2} is similar to that of Proposition~\ref{prop:inj} above, and will therefore be omitted.

\section{Proof of Theorem \ref{th2}}\label{proofTh1}
Since $R>R_{\text{B}}(\varepsilon)$ and $k=\lfloor Rn \rfloor$, we have
\ban a_n(\varepsilon)\overset{\textbf{.}}{<} \frac{k}{n}, \label{eq:start}\ean
which, together with the definition of $a_n(\varepsilon)$, implies that there exists a sequence\footnote{The definition of $\us$ is to be understood in the sense that the sequence index $n$ is dropped for simplicity of exposition.} $\us := \us_n\subset\IR^n$ such that 
\begin{align}
\MIL(\us) &\overset{\textbf{.}}{<} k \label{eq:suff1}\\ 
\pr[\xrv\in \us] &\overset{\textbf{.}}{\geqslant} 1-\varepsilon. \label{eq:suff2}
\end{align} 
For the remainder of the proof we choose $n$ to be sufficiently large for \eqref{eq:suff1} and \eqref{eq:suff2} to hold in the ${\#}$-sense. For $\am\in\IR^{k\times (n-\ell)}$ and $\bm\in\IR^{k\times \ell}$
define the separator\footnote{Taking ``error'' to be an arbitrary element of $\mathbb R^n\setminus \us$ we obtain a  measurable map $g\colon \IR^k\to \IR^{n-\ell}\times \IR^\ell$ as required in Definition~\ref{defsourcerate}.} 
\begin{align}\label{eq:decoder}
g(\vv)=
\begin{cases}
\xv,&\quad \text{if}\ \{\uv \mid [\am \:\: \bm]\uv= \vv\} \cap \us=\{\xv\}\\
%[\am,\bm]^{-1}(\vv) \cap \us,&\quad \text{if}\ [\am,\bm]^{-1}(\vv) \cap \us=\{ \x\}\\
\text{error},&\quad \text{else}.
\end{cases}
\end{align}
Then 
\begin{align}
&\pr [g([\am\:\: \bm]\xrv)\neq\xrv]\nonumber\\
&\overset{\phantom{(a)}}{=}\pr [g([\am\:\: \bm]\xrv)\neq\xrv, \xrv\in\us]\nonumber  \\ &\quad \quad\quad \quad\quad +\pr [g([\am\:\:\bm]\xrv)\neq\xrv, \xrv\notin\us]\label{eq:dec} \\ \nonumber
%&\leqslant \pr [g([\am,\bm]\xrv)\neq\xrv, \xrv\in\us]+\pr [\xrv\notin\us]\nonumber\\
&\overset{(a)}{\leqslant} \pr [g([\am\:\: \bm]\xrv)= \text{error}, \xrv\in\us]+\varepsilon \nonumber\\
&\overset{(b)}{\leqslant} \,\pr [\exists \uv \in \us_{\xrv}\!\setminus\! \{\0v\}: [\am\:\: \bm] \uv= \0v, \xrv\in\us]+\varepsilon , \nonumber
\end{align}
where $\us_{\xrv}=\us-\{\xrv\}=\{\uv - \xrv \mid \uv\in \us\}$, $(a)$ follows from the definition of the separator \eqref{eq:decoder} and from \eqref{eq:suff2}, and $(b)$ again is by  definition of the separator \eqref{eq:decoder}. Since $\MIL(\us)=\MIL(\us_{\xv})$, we find, through application of Proposition~\ref{prop:inj2}, that
\begin{align}\label{eq:pr2}
\leb^{k(n-\ell)}\{\am \mid  \exists \uv \in \us_{\xv}\!\setminus\! \{\0v\} : [\am\:\: \bm]\uv =\0v \}=0,
\end{align}
for all $\xv$. Therefore, the integral of \eqref{eq:pr2} with respect to $\mu_{\xrv}(d\xv)$ is zero, and, noting that \eqref{eq:pr2} can be written as an integral with respect to $d\am$, we can apply Fubini's Theorem to interchange the two integrals and obtain
\ban 	\int_{\mathbb R^{k\times (n-\ell)}}\pr [\exists \uv \in \us_{\xrv}\!\setminus\! \{\0v\}: [\am\:\: \bm] \uv= \0v, \xrv\in\us]d\am=0.	\label{eq:fub}	\ean
Therefore, we have $\pr [\exists \uv \in \us_{\xrv}\!\setminus \! \{\0v\}: [\am\:\: \bm]\uv= \0v, \xrv\in\us]=0$ for a.a.\ $\am\in\mathbb R^{k\times (n-\ell)}$.
In summary, we have shown that
\ba 	\pr [g([\am\:\: \bm]\xrv)\neq\xrv]\leqslant \varepsilon,		\ea
for a.a.\ $\am$, which completes the proof. 
\endproof

\section{Simplifying the proof of \cite[Thm.~18, 1)]{WV10}}
\label{sec:simple}

In this section, we sketch how the technique developed in the proof of Proposition~\ref{prop:inj} can be applied to devise a simplified and elementary proof of \cite[Thm.~18, 1)]{WV10}. The framework of almost lossless analog compression in \cite{WV10} for the case of  linear measurements and  a measurable decoder considers a general stochastic source process $\xrv$. The problem is to reconstruct $\xrv$ from $\hm \xrv$, where $\hm$ is the measurement matrix. The result in \cite[Thm.~18, 1)]{WV10} says that for $R>\! R_\text{B}(\varepsilon)$, for a.a.\ $\hm\in\mathbb R^{k\times n}$, there exists a measurable decoder $g$ such that
\ba \pr[g(\hm \xrv)\neq \xrv]\overset{\textbf{.}}{<}\varepsilon, 	\ea
where $k=\lfloor Rn\rfloor$.

Using  Proposition~\ref{prop:inj}, we can give an alternative, simplified proof of this result as follows. We choose a set $\us\subset \IR^n$ such that \eqref{eq:suff1} and \eqref{eq:suff2} hold, and define the decoder according to \begin{align}\label{eq:decoder2}
g(\vv)=
\begin{cases}
\xv,&\quad \text{if}\ \{\uv \mid \hm\uv= \vv\} \cap \us=\{\xv\}\\
%[\am,\bm]^{-1}(\vv) \cap \us,&\quad \text{if}\ [\am,\bm]^{-1}(\vv) \cap \us=\{ \x\}\\
\text{error},&\quad \text{else}.
\end{cases}
\end{align}
The probability of a decoding error is then decomposed as in \eqref{eq:dec}.  Applying Proposition~\ref{prop:inj} we find that a.a.\ matrices $\hm$ are injective on $\us$. Finally, invoking Fubini's Theorem as in the argument leading to \eqref{eq:fub} allows us to conclude that the probability of  decoding error  is zero when $\xrv\in \us$, leaving the total probability of  decoding error to be smaller than $\varepsilon$ and thus finishing the proof. 

\section{Sketch of the Proof of Lemma \ref{lemmax}}\label{prooflemmax}
Recall the role of $\lambda$ in Definition~\ref{definitionX}. The cases  $\lambda\in\{0,1\}$ are equivalent to the case $\lambda=1/2$, $\rho_1=\rho_2$, $\mu_{\text{d}_1}=\mu_{\text{d}_2}$, and $\mu_{\text{c}_1}=\mu_{\text{c}_2}$. 
 Hence, we can assume, without loss of generality,  that $0<\lambda<1$. This implies that we can take  $\ell=\lfloor\lambda n\rfloor \overset{\textbf{.}}{\notin} \{0,n\}$. 

Let $\as_i$ be the set of atoms of $\mu_{\text{d}_i}$. Then 
\begin{align*}
\ee[\1v_{X_j\notin\as_i}]
&=\mu_{\IrX_j}(\as_i^c)\\
&=
\begin{cases}
\rho_1,\quad \text{for}\ i=1,\ j\in\{1,\dots,n-\ell\}\\
\rho_2,\quad \text{for}\ i=2,\ j\in\{n-\ell+1,\dots,n\}.
\end{cases}
\end{align*}
By the weak law of large numbers, 
\begin{align*}
\frac{1}{n-\ell}\sum_{j=1}^{n-\ell} \1v_{X_j\notin\as_1} &\overset{\pr}{\to} \rho_1\nonumber\\
\frac{1}{\ell}\sum_{j=n-\ell+1}^n \1v_{X_j\notin\as_2} &\overset{\pr}{\to} \rho_2 ,
\end{align*}
which yields
\begin{align}\label{eq:wlln}
\frac{|\spt (\xrv)|}{n}
&=\frac{n-\ell}{n}\frac{1}{n-\ell}\sum_{j=1}^{n-\ell} \1v_{X_j\notin\as_1}+ \frac{\ell}{n}\frac{1}{\ell} \sum_{j=n-\ell+1}^n \1v_{X_j\notin\as_2}\nonumber\\
&\overset{\pr}{\to}(1-\lambda)\rho_1 + \lambda\rho_2
\end{align}
with the generalized support 
\begin{align*}%\label{eq:gensupport}
\spt(\xv)
&=\{i\in\{1,\dots,n-\ell \}\mid x_i\notin\as_1\}\nonumber\\
&\phantom{=\ }\cup \{i\in\{n-\ell+1,\dots,n\}\mid x_i\notin\as_2\}.
\end{align*}
Let $\kappa >0$ be arbitrary and set  
\begin{align*}
\cs&:=\{\xv\mid |\spt(\xv)|< ((1-\lambda)\rho_1 + \lambda\rho_2+\kappa)n\}\\
\ds&:=\{\xv\mid |\spt(\xv)|> ((1-\lambda)\rho_1 + \lambda\rho_2-\kappa)n\}. 
\end{align*}
Then, by convergence in probability in \eqref{eq:wlln}, we have 
\begin{align}
\pr[\xrv\in\cs]&\overset{\textbf{.}}{\geqslant}1-\varepsilon\label{eq:ass1lemma}\\
\pr[\xrv\in\ds]&\overset{\textbf{.}}{\geqslant}1-\varepsilon.\label{eq:ass2lemma}
\end{align}

The remaining steps of the proof are almost identical to the proof of \cite[Thm.~15]{WV10} and are therefore omitted. The idea is to decompose $\cs$ and $\ds$ into basic subsets, whose elements have certain components equal elements of the atomic sets $\as_{1},\as_{2}$, and the remaining components arbitrary. This allows us to bound the Minkowski dimension of $\cs$ and $\ts\cap\ds$, for arbitrary $\ts$ with  $\pr[\xrv\in\ts]\geqslant1-\varepsilon$, and thus to sandwich the Minkowski dimension compression rate according to
\ba  (1-\lambda)\rho_1 + \lambda\rho_2-\kappa \leqslant R_{\text{B}}(\varepsilon) \leqslant (1-\lambda)\rho_1 + \lambda\rho_2+\kappa, \ea
which yields the claim, since $\kappa$ is arbitrary.

\bibliographystyle{IEEEtran}
\bibliography{IEEEabrv,refs}
\end{document}